\documentclass[prc,floatfix,groupedaddress,nofootinbib,showpacs,preprint,amsmath,amssymb,amsfonts,superscriptaddress,widetable]{revtex4-1}

\usepackage{graphicx}
\usepackage[T1]{fontenc}
\begin{document}

\title{Rotational properties of the odd-$Z$ transfermium nucleus $^{255}$Lr by a particle-number-conserving method in the cranked shell model}

\author{Yu-Chun Li}%
\email{}
 \affiliation{College of Material Science and Technology, Nanjing University of
              Aeronautics and Astronautics, Nanjing 210016, China;}
\author{Xiao-Tao He}%
\email{Corresponding author (email: hext@nuaa.edu.cn)}
 \affiliation{College of Material Science and Technology, Nanjing University of
              Aeronautics and Astronautics, Nanjing 210016, China\\
              Received February 5, 2016; accepted February 29, 2016}

\date{\today}

\begin{abstract}
Experimentally observed ground state band based on the $1/2^{-}[521]$ Nilsson state and the first exited band based on the $7/2^{-}[514]$ Nilsson state of the odd-$Z$ nucleus $^{255}$Lr are studied by the cranked shell model (CSM) with the paring correlations treated by the particle-number-conserving (PNC) method. This is the first time the detailed theoretical investigations are performed on these rotational bands. Both experimental kinematic and dynamic moments of inertia ($\mathcal{J}^{(1)}$ and $\mathcal{J}^{(2)}$) versus rotational frequency are reproduced quite well by the PNC-CSM calculations. By comparing the theoretical kinematic moment of inertia $\mathcal{J}^{(1)}$ with the experimental ones extracted from different spin assignments, the spin $17/2^{-}\rightarrow13/2^{-}$ is assigned to the lowest-lying $196.6(5)$ keV transition of the $1/2^{-}[521]$ band, and $15/2^{-}\rightarrow11/2^{-}$ to the $189(1)$ keV transition of the $7/2^{-}[514]$ band, respectively. The proton $N=7$ major shell is included in the calculations. The intruder of the high$-j$ low$-\Omega$ $1j_{15/2}$ $ (1/2^{-}[770])$ orbital at the high spin leads to band-crossings at $\hbar\omega\approx0.20$ ($\hbar\omega\approx0.25$) MeV for the $7/2^{-}[514]$ $\alpha=-1/2$ ($\alpha=+1/2$) band, and at $\hbar\omega\approx0.175$ MeV for the $1/2^{-}[521]$ $\alpha=-1/2$ band, respectively. Further investigations show that the band-crossing frequencies are quadrupole deformation dependent.
\end{abstract}


%\keywords{Rotational band, Spin assignment, Band-crossing, High$-j$ orbital, Transfermium nuclei}%U

\maketitle

\section{Introduction}

There is an increasing interest in the detailed spectroscopic study of the nuclei in the transfermium mass region ($Z\sim100$) from the very beginning of this century (see reviews refs.~\cite{Leino2004_review,Herzberg2004_review,Herzberg2008_review} and references therein). Now rich experimental spectroscopic data have been reported in these nuclei for both of the ground and isomeric states, which provide useful informations to test and constrain theoretical models. Among these, the highest-$Z$ isotopes in which the rotational bands being observed is rutherfordium~\cite{Qian2009_257Rf,Greenlees2012_256Rf,Rissanen2013_257Rf}.  Besides the significance to understand the structure and reaction properties of these heavy nuclei, there is a hope that the single-particle studies of the transfermium nuclei could lead to a more reliable prediction of the location of the island of spherical superheavy nuclei. A particular attentions have been paid to the odd-mass nuclei. Part of the attraction comes from the fact that odd-mass nuclei can provide an additional fingerprint through the Nilsson configuration assignment to the rotational band. While the rotational spectroscopic data of odd-neutron transfermium nuclei are increasing slowly (taking refs.~\cite{Herzberg2002_No253,Reiter2005_No253,Ahmad2005_Cf251,Qian2009_257Rf,Herzberg2009_No253,Tandel2010_CmCf,Hessberger2012_No253,Rissanen2013_257Rf} for example.), observations in odd-proton transfermium nuclei are rare. Particularly, the rotational bands up to high spin have been observed only in $^{251}$Md and $^{255}$Lr, namely the one-quasiparticle band built on the $1/2^{-}[521]$ Nilsson state (configuration assigned by comparing to Hartree-Fock-Bogolyubov (HFB) calculations) in $^{251}$Md~\cite{Chatillon2007_Md251} and two bands in $^{255}$Lr which were tentatively assigned to be based on the $1/2^{-}[521]$ and $7/2^{-}[514]$ Nilsson states~\cite{Ketelhut2009_Lr255}. In addition, there are two reports of  the discovery of the high-$K$ isomeric states in $^{255}$Lr in refs.~\cite{Hauschild2008_Lr255_highK,Jeppesen09_Lr255_HighK}. The proton $1/2^{-}[521]$ orbital is of particular interest since it stems from the spherical $2 f_{5/2}$ orbital. The spin-orbit interaction strength of $2 f_{5/2}$ - $2 f_{7/2}$ partner governs the size of the possible $Z = 114$ spherical shell gap. Based on the transition energies calculated with HFB, the spin of the lowest observed transition at $195.4(3)$ keV of the $1/2^{-}[521]$ band in $^{251}$Md is assigned as $17/2\rightarrow13/2$. The spin assignment of the two bands based on $1/2^{-}[521]$ and $7/2^{-}[514]$ states in $^{255}$Lr is still open. Further theoretical investigation would be required to address this problem.

Systematic theoretical investigations have been performed both within the microscopic-macroscopic models~\cite{Cwiok1994,Muntian1999,Sobiczewski2001,FuXM2014_N150, LiuHL2012_No, LiuHL2011_high-order, ZhangZH2011,ZhangZH2012,XieKP2014} and self-consistent approaches~\cite{Bender2003_RMP,Cwiok1996,Cwiok1999,Duguet2001,Bender2003,Egido2000,Delaroche2006,Warda2012,Afanasjev2003,Vretenar2005,Litvinova2012,Prassa2012,Afanasjev2013}. Comparing to the constantly emerging theoretical investigations on the even-even nuclei, detailed studies of the properties of odd-mass nuclei only appear occasionally. The situation is even worse for odd-proton nuclei on which only a few theoretical investigations~\cite{He09_odd_mass,Afanasjev2013,Shirikova2013_Z=100,Adamian2010_OddZ} performed so far. Ref.~\cite{He09_odd_mass} is the first PNC-CSM study of the transfermium nuclei, in which the proton $N=7$ major shell was included to discuss the single-particle and rotational properties of the odd-neutron $^{253}$No and odd-proton $^{251}$Md nuclei. This enables to discuss the impact of the high-$j$ intruder proton $1j_{15/2}$ orbital on the rotational properties. The investigations lead to the conclusion that there is a considerable effect of the proton $1j_{15/2}(1/2^-[770])$ orbital on the rotational properties in transfermium nuclei at the high spin region. This study is one of the only (to our best knowledge) investigation performed with including of the contributions from the proton $N=7$ major shell in this mass region so far. However, the position of the $1j_{15/2}(1/2^-[770])$ orbital is very sensitive to the quadrupole deformation~\cite{Chasman1977}. Referring to the experimental deduced deformations in the neighboring $^{250}$Fm and $^{252,254}$No nuclei~\cite{Reiter1999_No254,Leino1999_No254,Bastin2006_Fm250,Herzberg2001_No252}, the quadrupole deformations used in ref.~\cite{He09_odd_mass} are $\varepsilon_{2}=0.30$ and $\varepsilon_{2}=0.29$ for $^{251}$Md and $^{253}$No, respectively, which are larger than the values used in almost all the other theoretical investigations. Please see sect.~\ref{sec:parameters} for the present deformation parameters situation in this mass region. Now under the condition that numerous theoretical studies aimed at modelling the experimental data so as to make the calculations of the heavier nuclei as reliable as possible, especially with the new set of Nilsson parameters $\kappa$ and $\mu$ obtained by fitting the experimental band head energies in more than 30 odd$-A$ nuclei with $Z\sim100$ within the frame work of PNC-CSM method~\cite{ZhangZH2011,ZhangZH2012}, to study the rotational properties of heavier nuclei, like $^{255}$Lr and to check the effect of the high-$j$ orbital and its deformation dependence are essential.

In the present work, PNC-CSM method is used to study the single-particle and rotational properties of the odd-proton nucleus $^{255}$Lr. The PNC-CSM method is proposed to treat properly the pairing correlations and blocking effects. It has been applied successfully for describing the properties of normal deformed nuclei in $A\sim 170$ mass region~\cite{Zeng1994_CMPC,WuCS1991_Yb,Zeng2002_Hf,LiuSX2002_Hf_nonadditivity,LiuSX2004_Yb,ZengJY2001_rare_earth}, superdeformed nuclei in $A\sim 150,190$ mass region~\cite{ZengJY1991_A150_Kstructure,LiuSX2000_A150_spin,WuCS1992_A190,LiuSX2002_Hg,LiuSX2004_Hg193,HeXT2005_Tl,HeXT2005_Tl_IB}, high-K isomeric states in the rare-earth and actinide mass region~\cite{ZhangZH2009_NPA,ZhangZH2009_PRC,ZhangZH2010_CPC,ZhangZH2010_CPC_b,FuXM2013,FuXM2013_PRC,LiBH2013} and recently in the heaviest actinides and light superheavy nuclei around $Z\sim 100$ region~\cite{ZhangZH2011,ZhangZH2012,He09_odd_mass,ZhangZH2013_256Rf}. In contrast to the conventional Bardeen-Cooper-Schrieffer (BCS) or HFB approach, the Hamiltonian is diagonalized directly in a truncated Fock space in the PNC method~\cite{Zeng1983,Zeng1994b}. Therefore, particle number is conserved and Pauli blocking effects are taken into account exactly.

\section{Theoretical formalism}\label{sec:2}

The details of the PNC-CSM method can be found in refs.~\cite{Zeng1983,Zeng1994_CMPC,Zeng1994b}. For convenience, we give briefly the related formalism here. The cranked shell model hamiltonian of an axially symmetric nucleus in the rotating frame is expressed as:
\begin{equation}
\ H_{\text{CSM}}=H_{0}+H_{\text{P}}=\sum_{n}(h_{\text{Nil}}-\omega j_{x})_{n}+H_{\text{P}}(0)+H_{\text{P}}(2)\ ,
\end{equation}
where $h_{\textrm{Nil}}$ is the Nilsson Hamiltonian \cite{Nilsson1955,Nilsson1969},
$-\omega j_{x}$ is the Coriolis force with the cranking frequency $
\omega$ about the $x$ axis (perpendicular to the nuclear symmetry $z$ axis). $H_{\text{P}}$ is the pairing including monopole and quadrupole pairing correlations,
\begin{eqnarray}
 \ H_{\text{P}}(0)
 =-G_{0}\sum_{\xi \eta }
 a_{\xi }^{\dagger}a_{\overline{\xi }}^{\dagger }a_{\overline{\eta }}a_{\eta }\ ,
 \\
 H_{\text{P}}(2)
 =-G_{2}\sum_{\xi \eta } q_{2}(\xi) q_{2}(\eta)
 a_{\xi}^{\dagger } a_{\overline{\xi}}^{\dagger}
 a_{\overline{\eta}}a_{\eta}\ ,
 \label{eq:Hp}
\end{eqnarray}
with $\overline{\xi }$ and $\overline{\eta }$ being the time-reversal
states of the Nilsson state $\xi$ and $\eta$, respectively. The quantity $q_{2}(\xi) =\sqrt{16\pi /5}\left\langle \xi \right\vert r^{2}Y_{20}\left\vert \xi \right\rangle $ is the
diagonal element of the stretched quadrupole operator, and $G_{0}$ and $G_{2}$ are the effective
strengths of monopole and quadrupole pairing interactions, respectively.

In the PNC-CSM calculation, $h_{0}(\omega)=h_{\textrm{Nil}}-\omega j_{x}$ is diagonalized firstly to obtain the cranked Nilsson orbitals (see Figure~\ref{fig:Fig1}). Then $H_{\text{CSM}}$ is diagonalized in a
sufficiently large Cranked Many-Particle Configuration (CMPC) space
to obtain the yrast and low-lying eigenstates. Instead of
the usual single-particle level truncation in
common shell-model calculations, a cranked many-particle
configuration truncation (Fock space truncation)
is adopted, which is crucial to make the PNC calculations
for low-lying excited states both workable and sufficiently
accurate~\cite{WuCS1989} . The eigenstate of $H_{\text{CSM}}$ is expressed as:
\begin{equation}
\left| \psi \right\rangle =\sum_{i}C_{i}\left| i\right\rangle \ ,
\label{eq:wf}
\end{equation}
where $\left\vert i\right\rangle $ is a cranked many-particle configuration
(an occupation of particles in the cranked Nilsson orbitals) and $C_{i}$ is the
corresponding probability amplitude.

The angular momentum alignment $\left\langle J_{x} \right\rangle$ of
the state $\left\vert \psi \right\rangle$ is given by
\begin{equation}
 \left\langle \psi \right| J_{x}
 \left| \psi \right\rangle
 = \sum_{i}\left|C_{i}\right| ^{2}
   \left\langle i\right| J_{x}\left| i\right\rangle
 + 2\sum_{i<j}C_{i}^{\ast }C_{j}
   \left\langle i\right| J_{x}\left| j\right\rangle\ .
 \label{eq:Jx}
\end{equation}
The kinematic moment of inertia (MOI) is $\mathcal{J}^{(1)}=\left\langle \psi \right\vert J_{x}\left\vert\psi \right\rangle /\omega $, and the dynamical moment of inertia is $\mathcal{J}^{(2)}={\rm d}\left\langle \psi \right\vert J_{x}\left\vert\psi \right\rangle /{\rm d}\omega $.

\section{Parameters}\label{sec:parameters}

The deformations are input parameters in the PNC-CSM calculations. Normally the quadrupole deformation parameter $\varepsilon_{2}$ is chosen to be consistent with the value deduced by experiment. In the transfermium nuclei mass region there are only a few experimental reports of the quadrupole deformation, namely $\beta_{2}=0.27\pm0.02$ for $^{254}$No in refs.~\cite{Reiter1999_No254,Leino1999_No254}, $\beta_{2}=0.28\pm0.02$ for $^{250}$Fm in ref.~\cite{Bastin2006_Fm250},  $\beta_{2}=0.31\pm 0.02$ for $^{252}$No and $\beta_{2}=0.32\pm0.02$ for $^{254}$No in ref.~\cite{Herzberg2001_No252}. They are not consistent. The values predicted by different theoretical models are various too (taking refs.~\cite{Cwiok1994,Sobiczewski2001,Mueller1995_nucleidata} for example). At the present stage almost all theoretical calculations predicted (or used) smaller quadrupole deformations when comparing with the experimental values shown above. Since the experimental data are not enough yet, it is still too early to answer the question that whether the deformations in theoretical studies for transfermium nuclei are underestimated or not. In the present work, $\varepsilon_{2}=0.27$ and $\varepsilon_{4}=0.02$ are accepted by changing the values smoothly along the $N=152$ isotone in Table II of ref.~\cite{ZhangZH2012}. It has been noted that $\beta_{6}$ deformation can be significant~\cite{LiuHL2011_high-order,Muntian2001_deformation}  and can have a measurable effect on the structure of the nuclei in this mass region~\cite{LiuHL2012_No,ZhangZH2013_256Rf}. We include $\varepsilon_{6}=0.02$ in the present  calculations, which is closed to the value in configuration-constrained potential-energy-surface (PES) calculation in ref.~\cite{LiuHL2011_high-order}.

The new set of Nilsson parameters $(\kappa,\mu)$, which optimized to reproduce the experimental level schemes for the transfermium nuclei in refs.~\cite{ZhangZH2011,ZhangZH2012} are used in this work.  The values of proton $\kappa_{5},\mu_{5}$ are modified slightly to reproduce the correct single-particle level sequence when $\varepsilon_{6}$ is included. In addition to the optimized $(\kappa,\mu)$ in refs.~\cite{ZhangZH2011,ZhangZH2012},  proton $\kappa_{7}=0.057$ and $\mu_{7}=0.654$ are adopted in this work.

The effective pairing strengths $G_{0}$ and $G_{2}$ can be determined by the odd-even differences in nuclear binding energies. They are connected with the dimension of the truncated CMPC space. The CMPC space for $^{255}$Lr is constructed in the proton $N$ = 4, 5, 6, 7 shells and the neutron $N$ = 6, 7 shells. The dimensions of the CMPC space for $^{255}$Lr are about 1000 for both of protons and neutrons. The corresponding effective monopole and quadrupole pairing strengths are
$G_{0p} = 0.25$ MeV, $G_{2p} = 0.01$ MeV, $G_{0n} = 0.25$ MeV, and $G_{2n} = 0.02$ MeV~\cite{ZhangZH2012}. As we are only interested in the yrast and low-lying excited states, the number of the important CMPC's involved (weight $>1\%$) is not very large (usually $<20$) and almost all the CMPC's with weight $ >0.1\%$ are included in.

\section{Results and discussions}\label{sec:4}

\begin{figure}[h]
\centering
    \includegraphics[width=6cm]{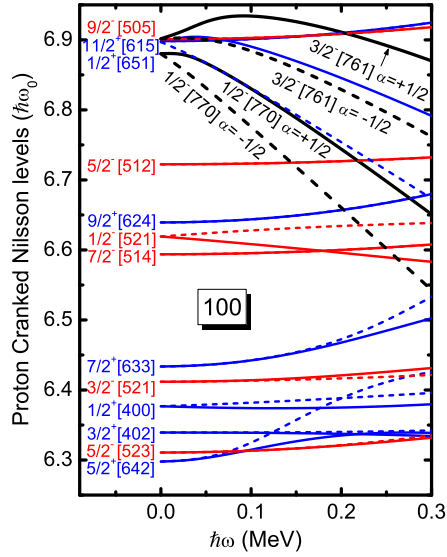}
    \caption{(Color online) Proton cranked Nilsson levels near the Fermi surface of $^{255}$Lr. The signature $\alpha=+1/2$ ($\alpha=-1/2$) is denoted by solid (dash) lines. The positive (negative) parity is denoted by blue (red) lines. The high-$j$ orbitals $1/2^{-}[770]$ and $3/2^{-}[761]$ are denoted by black lines. The deformation parameters are $\varepsilon_{2}=0.27$, $\varepsilon_{4}=0.02$ and $\varepsilon_{6}=0.02$.} \label{fig:Fig1}
\end{figure}

Figure\ref{fig:Fig1} shows the proton cranked Nilsson levels near the Fermi surface of $^{255}$Lr. The signature $\alpha=+1/2$ ($\alpha=-1/2$) is denoted by solid (dash) lines. The positive (negative) parity is denoted by blue (red) lines. The high-$j$ orbitals $1/2^{-}[770]$ and $3/2^{-}[761]$ are denoted by black lines. It is seen that the high-$j$ orbital $1j_{15/2}(1/2^{-}[770])$ slopes down sharply with increasing rotational frequency $\hbar\omega$ and the $\alpha=-1/2$ one crosses with $1/2^{-}[521]$ $\alpha=-1/2$ orbital at $\hbar\omega\approx0.20$ MeV.

%%%%%%%%%%%%%%%%%%%%%%%%%%%%%%%%%%%%%%%%%%%%%%%%%%%%%
%%%%%%%%%%%%%%%%%%%%%%%%%%%%%%%%%%%%%%%%%%%%%%%%%%%%%

The ground state and the first exited state of $^{255}$Lr have been determined through $\alpha$ decay properties to be $1/2^{-}[521]$ and $7/2^{-}[514]$, respectively~\cite{Chatillon2006_OddZ}. The properties of $1/2^{-}[521]$ orbital were discussed by Ahmad et al.~\cite{Ahmad1977_521} where the decoupling parameter was stated to be closed to 1. For such a decoupling parameter of the $K=1/2$ band, the nonyrast sequence is almost degenerate with the yrast sequence as shown in the $^{251}$Md~\cite{Chatillon2007_Md251}. This leads to the decay proceed mainly through $E2$ transitions in the $\alpha=+1/2$ band. The decay patterns in $^{255}$Lr are expected to be similar to that in $^{251}$Md. The sequence of $\gamma$ rays [196.6(5), 247.2(5), 296.2(5), 342.9(5), 387(1) and 430(1) keV] observed by Ketelhut et al.~\cite{Ketelhut2009_Lr255} was assigned to be a rotational band based on the $1/2^{-}[521]$ state in $^{255}$Lr. There is no spin assignment for these transitions. Note that the spin of the lowest-lying $195.4(3)$ keV transition of the $1/2^{-}[521]$ band in $^{251}$Md is tentatively assigned to be  $17/2^{-}\rightarrow13/2^{-}$ (Figure 2 in ref.~\cite{Chatillon2007_Md251}). Another sequence of transitions with energies 189(1), 239(1), 288.4(5), 338(1), 384(1) and 215(1), 264.6(5), 314.0(5), 360(1) keV was observed at the mean time, which is tentatively assigned to be $E2$ transitions in both signatures of a strongly-coupled rotational band based on the $7/2^{-}[514]$ configuration in $^{255}$Lr. The spin assignment is absent too.

%%%%%%%%%%%%%%%%%%%%%%%%%%%%%%%%%%%%%%%%%%%%%%%%%%%%%
\begin{figure}[h]
\centering
    \includegraphics[width=8cm]{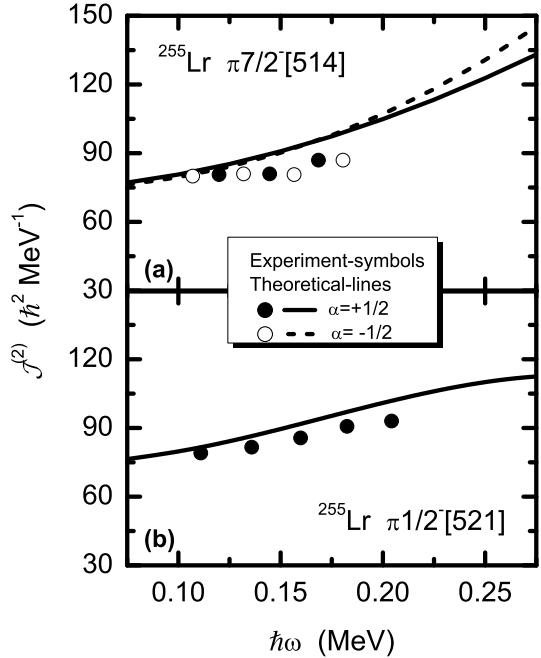}
    \caption{Theoretical and experimental dynamic moment of inertia $\mathcal{J}^{(2)}$ versus rotational frequency of the $K^\pi=7/2^-$ $(\alpha=\pm1/2)$ (a) and the $K^\pi=1/2^-$ $(\alpha=+1/2)$ (b) bands in $^{255}$Lr. Solid (open) circles denote the observed experimental $\alpha=+1/2$ ($\alpha=-1/2$) bands. Solid (dash) lines denote the calculated $\alpha=+1/2$ ($\alpha=-1/2$) bands. The experimental data are taken from ref.~\cite{Ketelhut2009_Lr255}.} \label{fig:Fig2}
\end{figure}
%%%%%%%%%%%%%%%%%%%%%%%%%%%%%%%%%%%%%%%%%%%%%%%%%%%%%

\begin{figure}[h]
\centering
    \includegraphics[width=8cm]{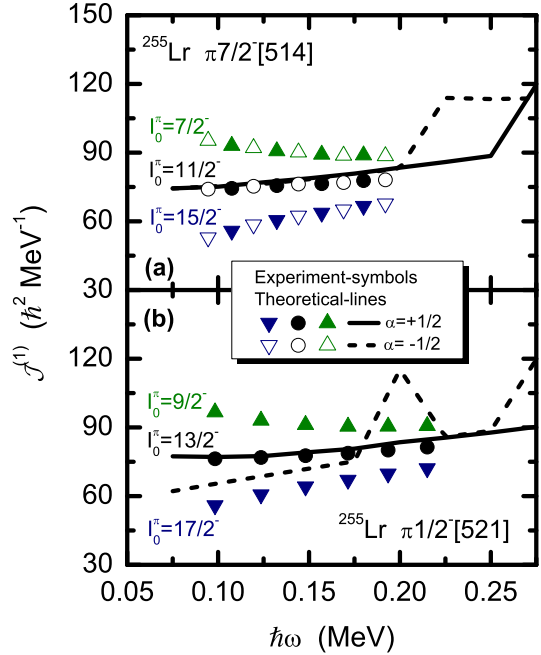}
    \caption{(Color online) Theoretical and experimental kinematic moment of inertia $\mathcal{J}^{(1)}$ versus rotational frequency of the $K^\pi=7/2^-$ (a) and $K^\pi=1/2^-$ (b) bands in $^{255}$Lr. (a) The dark green up-triangles, black circles and dark blue down-triangles denote the experimental $\mathcal{J}^{(1)}$ extracted by assigning the  lowest-lying $189(1)$  keV transition as $11/2^{-}\rightarrow7/2^{-}$ , $15/2^{-}\rightarrow11/2^{-}$  and $19/2^{-}\rightarrow15/2^{-}$, respectively. And (b) denote the experimental $\mathcal{J}^{(1)}$ by assigning the lowest-lying $196.6(5)$ keV transition as $13/2^{-}\rightarrow9/2^{-}$, $17/2^{-}\rightarrow13/2^{-}$ and $21/2^{-}\rightarrow17/2^{-}$, respectively. The experimental data are taken from ref.~\cite{Ketelhut2009_Lr255}.} \label{fig:Fig3}
\end{figure}

\begin{figure}[h]
\centering
    \includegraphics[width=10cm]{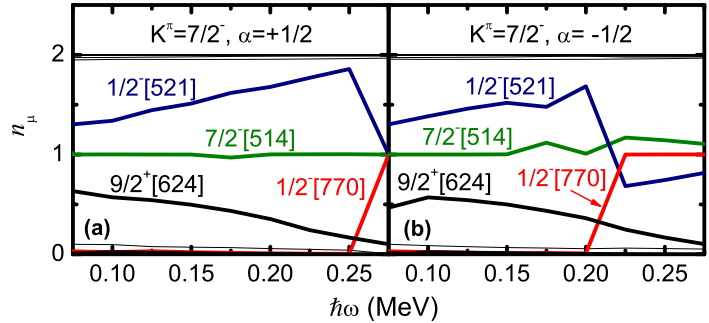}
    \caption{(Color online) Occupation probability $n_\mu$ of each cranked proton orbital $\mu$ (including both $\alpha=\pm1/2$) near the Fermi surface of $^{255}$Lr for the $K^\pi=7/2^-$ band with $\alpha=+1/2$ (a) and $\alpha=-1/2$ (b). Fully occupied ($n_{\mu}\approx2$) and empty ($n_{\mu}\approx0$) orbitals denoted by black thin lines are not labelled.} \label{fig:Fig4}
\end{figure}

The comparison of the theoretical $\mathcal{J}^{(2)}$ with the extracted experimental values for $1/2^{-}[521]$ and $7/2^{-}[514]$ configuration bands is plotted in Figure~\ref{fig:Fig2}. Note that the extracted dynamical moment of inertia $\mathcal{J}^{(2)}=2\Delta I/[E_{\gamma}(I)-E_{\gamma}(I-2)]$ is spin independent. For the sake of consistency, both signatures $\alpha=\pm1/2$ are plotted for the $7/2^{-}[514]$ band in Figure~\ref{fig:Fig2}(a), which according to the two sequences of $\gamma$ rays observed in experiment. The signature ($\alpha=I$ mod 2) assignment comes from the following spin assignment discussions (see Figure~\ref{fig:Fig3}). For $1/2^{-}[521]$, only $\alpha=+1/2$ signature-partner band is shown in Figure~\ref{fig:Fig2}(b) since the $E(2)$ transitions observed in experiment was assigned as $\alpha=+1/2$ band~\cite{Ketelhut2009_Lr255}. A fairly good agreement between the theoretical results and the experimental data achieves during the whole \textit{observed} rotational frequency for $1/2^{-}[521]$ $(\alpha=+1/2)$ and $7/2^{-}[514]$ $(\alpha=\pm1/2)$ bands. When it goes beyond the observed frequency region at $\hbar\omega>0.20$ MeV for $K^{\pi}=7/2^{-}$ band, a signature splitting appears in the calculation. Further investigations of the occupation probabilities $n_{\mu}$ of each proton single-particle orbitals near the Fermi surface of $^{255}$Lr (see Figure~\ref{fig:Fig4}) show that for the $\alpha=+1/2$ partner band,  the blocking of the $7/2^{-}[514]$ $\alpha=+1/2$ orbital ($n_{\mu}\approx1$) keeps to $\hbar\omega\approx0.275$ MeV whereas for the $\alpha=-1/2$ partner band, a band-crossing occurs at $\hbar\omega\approx0.20$ MeV. This results in the signature splitting at $\hbar\omega>0.20$ MeV, which will be discussed in detail in the following study of kinematic moment of inertia $\mathcal{J}^{(1)}$.

%%%%%%%%%%%%%%%%%%%%%%%%%%%%%%%%%%%%%%%%%%%%%%%%%%%%%
%%%%%%%%%%%%%%%%%%%%%%%%%%%%%%%%%%%%%%%%%%%%%%%%%%%%%

Under the condition that the PNC-CSM calculations reproduce the experimental $\mathcal{J}^{(2)}$ well, we further determine the band head spin $I_{0}^{\pi}$ by $\mathcal{J}^{(1)}$ as shown in Figure~\ref{fig:Fig3}. The dark green up-triangles, black circles and dark blue down-triangles denote the experimental $\mathcal{J}^{(1)}$ of $K^{\pi}=7/2^{-}$  band extracted by assigning the observed lowest-lying $189(1)$  keV transition as $11/2^{-}\rightarrow7/2^{-}$ , $15/2^{-}\rightarrow11/2^{-}$  and $19/2^{-}\rightarrow15/2^{-}$, respectively. The PNC-CSM calculation agrees well with the $15/2^{-}\rightarrow11/2^{-}$ assignment [see Figure~\ref{fig:Fig3}(a)]. As for $K^{\pi}=1/2^{-}$ band, same symbols are used to denote the $\mathcal{J}^{(1)}$ by assigning the lowest-lying $196.6(5)$ keV transition as $13/2^{-}\rightarrow9/2^{-}$, $17/2^{-}\rightarrow13/2^{-}$ and $21/2^{-}\rightarrow17/2^{-}$, respectively. The calculation agrees with the assignment of $17/2^{-}\rightarrow13/2^{-}$ [see Figure~\ref{fig:Fig3}(b)]. Note that the spin of the lowest observed $195.4(3)$ keV transition of the $1/2^{-}[521]$ band in $^{251}$Md is assigned as $17/2\rightarrow13/2$ by the transition energies calculations with HFB~\cite{Chatillon2007_Md251}. The following discussions are all based on the spin assignments of $I_{0}^{\pi}=11/2^{-}$ for $K^{\pi}=7/2^{-}$ band and $I_{0}^{\pi}=13/2^{-}$ for $K^{\pi}=1/2^{-}$ band, respectively.

%%%%%%%%%%%%%%%%%%%%%%%%%%%%%%%%%%%%%%%%%%%%%%%%%%%%%
%%%%%%%%%%%%%%%%%%%%%%%%%%%%%%%%%%%%%%%%%%%%%%%%%%%%%

$^{255}$Lr is the heaviest odd-$Z$ nucleus in which the rotational bands have been observed in experiment so far. The lowest single-particle orbital of $N=7$ major shell is $1/2^{-}[770]$, which locates at about $0.25\hbar\omega_{0}$ above the Fermi surface of $^{255}$Lr (at $\hbar\omega=0.0$ MeV with $\varepsilon_{2}=0.27)$ (see Figure~\ref{fig:Fig1}). It is far away from the Fermi surface of $^{255}$Lr, which could be a major reason that the effect of the proton $N=7$ major shell on the rotational properties of the transfermium nuclei is neglected by most of the theoretical studies. Consuming of the computer time could be another reason. However, as rotational frequency increasing, the $1/2^{-}[770]$ orbital slopes down quickly and gets close to the Fermi surface at $\hbar\omega\approx0.20$ MeV (see Figure~\ref{fig:Fig1}). In the PNC-CSM calculations, the \textit{cranked} many-particle configuration spaces are adopted. The truncated configurations are adjusted according to the variation of the single particle levels with rotational frequency. This enables one to study the effect of the high$-j$ intruder orbitals varying with rotational frequency even though these high-$j$ orbitals are absent at the low rotational frequency~\cite{Zeng1994_CMPC}.

%%%%%%%%%%%%%%%%%%%%%%%%%%%%%%%%%%%%%%%%%%%%%%%%%%%%%
\begin{figure}[h]
\centering
    \includegraphics[width=8cm]{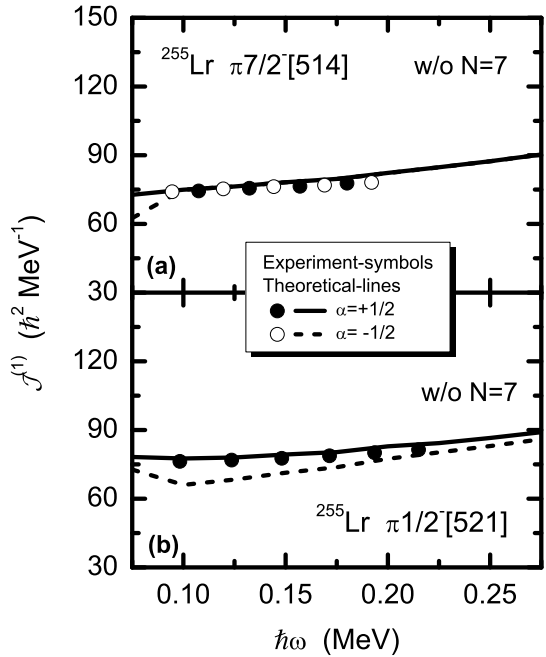}
    \caption{Kinematic moment of inertia $\mathcal{J}^{(1)}$ versus rotational frequency of the $K^\pi=7/2^-$ (a) and $K^\pi=1/2^-$ (b) bands calculated \textit{without} the proton $N=7$ major shell. Solid (dash) lines denote the theoretical $\mathcal{J}^{(1)}$ of $\alpha=+1/2$ $(\alpha=-1/2)$ band. The solid (open) circles denote the experimental $\mathcal{J}^{(1)}$ of $\alpha=+1/2$ $(\alpha=-1/2)$ band with the assigned spin $I_{0}^{\pi}=11/2^{-}$ for $K^{\pi}=7/2^{-}$ band and $I_{0}^{\pi}=13/2^{-}$ for $K^{\pi}=1/2^{-}$ band, respectively. } \label{fig:Fig5}
\end{figure}
%%%%%%%%%%%%%%%%%%%%%%%%%%%%%%%%%%%%%%%%%%

As shown in Figure~\ref{fig:Fig3}, the signature splitting of $\mathcal{J}^{(1)}$ of $K^{\pi}=7/2^{-}$ band is unambiguous. A sharp up-bending occurs at $\hbar\omega\approx0.20$ MeV for the $\alpha=-1/2$ band while it delays to $\hbar\omega\approx0.25$ MeV for the $\alpha=+1/2$ band. The intruder of the $1/2^{-}[770]$ orbital at high rotational frequency changes the occupation probabilities. For the $\alpha=+1/2$ band, the pure blocking ($n_{\mu}\approx1$) of $7/2^{-}[514]$ orbital is persisted up to $\hbar\omega\approx0.25$ MeV [see Figure~\ref{fig:Fig4}(a)] whereas the band-crossing between the one-quasiparticle band $7/2^{-}[514]$ and the three-quasiparticle band $7/2^{-}[514]\otimes1/2^{-}[521]\otimes1/2^{-}[770]$ occurs at $\hbar\omega>0.20$ MeV for the $\alpha=-1/2$ band [see Figure~\ref{fig:Fig4}(b)] . The band-crossing frequency difference ($\hbar\omega\approx0.20$ MeV for $\alpha=-1/2$ and $\hbar\omega\approx0.25$ MeV for $\alpha=+1/2$ band) is due to the signature splitting of the $1/2^{-}[770]$ orbital (see Figure~\ref{fig:Fig1}). As the high$-j$ low$-\Omega$ orbitals are characterized by their large contributions to alignment and Coriolis responses, increasing occupation probability of the $1/2^{-}[770]$ orbital leads to the sudden up-bending of the $\mathcal{J}^{(1)}$. For comparison, calculated $\mathcal{J}^{(1)}$ for $K^{\pi}=7/2^{-}$ band without including the proton $N=7$ major shell are plotted in Figure~\ref{fig:Fig5}(a) where both the up-bendings and the signature splitting at high rotational frequency are disappeared.

%%%%%%%%%%%%%%%%%%%%%%%%%%%%%%%%%%%%%%%%%%%%%%%%%%%%%
\begin{figure}[h]
\centering
    \includegraphics[width=10cm]{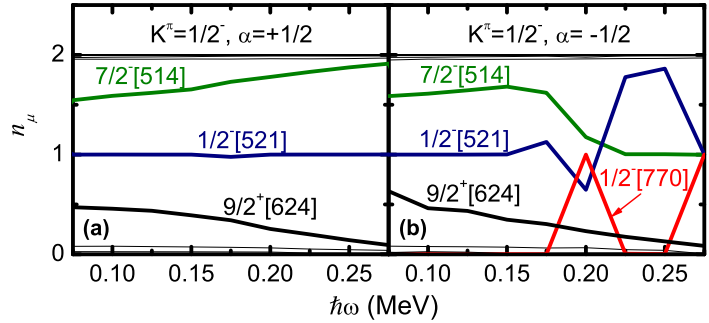}
    \caption{(Color online) Same as Figure~\ref{fig:Fig4}, but for $K^\pi=1/2^-$ band.} \label{fig:Fig6}
\end{figure}
%%%%%%%%%%%%%%%%%%%%%%%%%%%%%%%%%%%%%%%%%%%%%%%%%%%%%

As for $K^{\pi}=1/2^{-}$ band, due to the decoupling term, the $\alpha=-1/2$ signature band is pushed up in energy and only one signature-partner ($\alpha=+1/2$) band is observed in experiment. As is shown in Figure~\ref{fig:Fig3}(b), $\alpha=-1/2$ signature band is obtained in theoretical calculations as well as its signature-partner $\alpha=+1/2$ band. The behaviors of $\alpha=\pm1/2$ bands are very different, which is due to the signature splitting of both of $1/2^{-}[770]$ and $1/2^{-}[521]$ orbitals (see Figure~\ref{fig:Fig1}). The sudden upbending of $\mathcal{J}^{(1)}$ of $\alpha=-1/2$ band at $\hbar\omega\approx0.175$ MeV is mainly due to the alignment contributions from the $1/2^{-}[770]$ pairs. To analyze further by the occupation probability (see Figure~\ref{fig:Fig6}), the blocking of $1/2^{-}[521]$ keeps pure ($n_{\mu}\approx1$) at the whole calculated rotational frequency region for $\alpha=+1/2$ band whereas band-crossings occur at $\hbar\omega>0.175$ MeV for the $\alpha=-1/2$ band. As the rotational frequency increasing, resulted from the interplay between the $1/2^{-}[521]$, $7/2^{-}[514]$ and $1/2^{-}[770]$ orbitals, occupation of the $1/2^{-}[770]$ is not stable. It decreases from $n_{\mu}\approx1$ at $\hbar\omega\approx0.175$ MeV to $n_{\mu}\approx0$ at $\hbar\omega\approx0.225$-0.25 MeV, and increases again to $n_{\mu}\approx1$ at $\hbar\omega>0.275$ MeV.  The irregular behaviors of $\mathcal{J}^{(1)}$ for $\alpha=-1/2$ band at $\hbar\omega>0.175$ MeV result mainly from the effect of the $1/2^{-}[770]$ orbital. As the proton $N=7$ major shell does not included, the $\mathcal{J}^{(1)}$ goes smoothly with increasing rotational frequency [see Figure~\ref{fig:Fig5}(b)].

%%%%%%%%%%%%%%%%%%%%%%%%%%%%%%%%%%%%%%%%%%%%%%%%%%%%%%
\begin{figure}[h]
\centering
    \includegraphics[width=8cm]{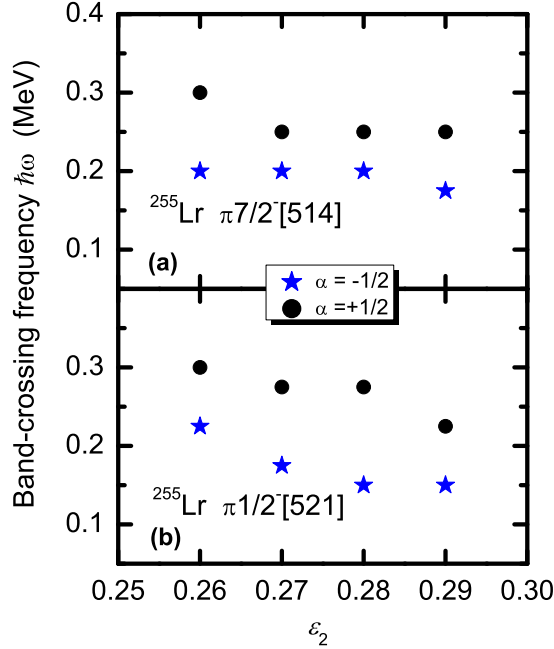}
    \caption{(Color online) Band-crossing frequency $\hbar\omega$ versus quadrupole deformation $\varepsilon_2$ of the $K^\pi=7/2^-$ (a) and $K^\pi=1/2^-$ (b) bands. The blue stars (black circles) denote the $\alpha=-1/2$ ($\alpha=+1/2$) band.}\label{fig:Fig7}
\end{figure}
%%%%%%%%%%%%%%%%%%%%%%%%%%%%%%%%%%%%%%%%%%%%%%%%%%%%%%

As the above discussions show that the $1/2^{-}[770]$ orbital could effect the high spin rotational properties a lot. Its position is crucial. The position of the $1/2^{-}[770]$ orbital is very sensitive to the deformation. The Woods-Saxon potential calculations by Chasman et al. display that the $1/2^{-}[770]$ orbital slopes down quickly with increasing quadrupole deformation (see Figure 4 in ref.~\cite{Chasman1977}). The band-crossing frequencies versus quadrupole deformation by PNC-CSM are plotted in Figure~\ref{fig:Fig7}. This provides us the informations that which frequency region in $^{255}$Lr could be effected by the $1/2^{-}[770]$ orbital under different quadrupole deformations. As the band-crossing frequencies of $\alpha=+1/2$ (black solid circles) and $\alpha=-1/2$ (blue stars) bands are different, the signature splittings are explicit for both of $K^{\pi}=7/2^{-}$ and $K^{\pi}=1/2^{-}$ bands. Because the $1/2^{-}[770]$ $\alpha=-1/2$ orbital goes down faster than its $\alpha=+1/2$ partner orbital (see Figure~\ref{fig:Fig1}), all the band-crossing frequencies for the $\alpha=-1/2$ bands are lower than that for its $\alpha=+1/2$ partner bands. The general trend shown in Figure~\ref{fig:Fig7} is a larger deformation leads to a lower band-crossing frequency. The $K^{\pi}=1/2^{-} (\alpha=-1/2)$ band is the most sensitive one to the variation of the deformations. The band-crossing frequency is $\hbar\omega\approx0.225$ MeV at $\varepsilon_{2}=0.26$, and decreases to $\hbar\omega\approx0.175$ MeV at $\varepsilon_{2}=0.27$, further to $\hbar\omega\approx0.150$ MeV at $\varepsilon_{2}=0.28,0.29$. For $K^{\pi}=7/2^{-}$ ($\alpha=-1/2$) band, except for $\varepsilon_{2}=0.29$ (the band-crossing frequency is $\hbar\omega\approx0.175$ MeV), the band-crossing frequencies keep constant at $\hbar\omega\approx0.20$ as varying the deformation from $\varepsilon_{2}=0.26$ to $\varepsilon_{2}=0.28$. As the effective frequency regions of $1/2^{-}[770]$ orbital are all beyond the nowadays experimentally observed frequency region, whether the $1/2^{-}[770]$ orbital plays an important role in transfermium nuclei needs further tests by future experiments.

\section{Summary}

Experimentally observed ground state band based on the $1/2^{-}[521]$ Nilsson state and the first exited band based on the $7/2^{-}[514]$ Nilsson state in the odd-$Z$ nucleus $^{255}$Lr are studied by the cranked shell model with the paring correlations treated by the particle-number-conserving method. To our best knowledges, this is the first time the detailed investigations are performed on these rotational bands. Both the experimental kinematic and dynamic moments of inertia versus rotational frequency are reproduced quite well by the PNC-CSM calculations. The spin of the lowest-lying $196.6(5)$ keV transition of the $1/2^{-}[521]$ band does not assigned experimentally, neither does the spin of the lowest-lying $189(1)$ keV transition of the $7/2^{-}[514]$ band. By comparing the theoretical kinematic moments of inertia with the experimental ones extracted from different spin assignments, the spin $17/2^{-}\rightarrow13/2^{-}$ is assigned to the lowest-lying $196.6(5)$ keV transition of the $1/2^{-}[521]$ band, and $15/2^{-}\rightarrow11/2^{-}$ to the $189(1)$ keV transition of the $7/2^{-}[514]$ band, respectively.

The proton $N=7$ major shell is included in the present calculations. Theoretical results predict a considerable effect of the high-$j$ low-$\Omega$ $1j_{15/2}$ $(1/2^{-}[770])$ orbital on the high spin behavior of these rotational bands. Due to the contributions from the $1j_{15/2}$ $(1/2^{-}[770])$ orbital, theoretical calculations predict band-crossings at $\hbar\omega\approx0.20$ ($\hbar\omega\approx0.25$) MeV for the $7/2^{-}[514]$ $\alpha=-1/2$ ($\alpha=+1/2$) band and at $\hbar\omega\approx0.175$ MeV for the $1/2^{-}[521]$ $\alpha=-1/2$ band, respectively. Since the position of $1/2^{-}[770]$ orbital is very sensitive to the deformation, these band-crossing frequencies are deformation dependent. A larger quadrupole deformation results in a lower band-crossing frequency. Because the $1/2^{-}[770]$ $\alpha=-1/2$ orbital goes down faster than its $\alpha=+1/2$ partner orbital, the band-crossing frequency versus quadrupole deformation is lower for the $\alpha=-1/2$ bands than that for its $\alpha=+1/2$ partner bands. Varying the deformation from $\varepsilon_{2}=0.26$ to $\varepsilon_{2}=0.29$, the lowest band-crossing frequency is at $\hbar\omega=0.175$ MeV for the $7/2^{-}[514]$ $\alpha=-1/2$ band, and at $\hbar\omega=0.15$ MeV for the $1/2^{-}[521]$ $\alpha=-1/2$ band, respectively. Since lack of enough experimental data for the odd-$Z$ transfermium nuclei, whether the $1j_{15/2}$ $1/2^{-}[770]$ orbital plays an important role in the rotational properties of transfermium nuclei needs further tests by future experiments.

\acknowledgements{ This work was supported by the National Natural Science Foundation of China (Grant Nos. 11275098 and 11275067) and the Priority Academic Program Development of Jiangsu Higher Education Institutions.}

\end{document}